\documentclass[aps,preprint,onecolumn,floatfix,superscriptaddress,nofootinbib,pra]{revtex4-1}

\usepackage{amsmath}
\usepackage{graphicx,bm}
\usepackage[colorlinks,citecolor=blue]{hyperref}
\usepackage[caption=false]{subfig}
\usepackage{slashed}
\usepackage{ulem}
\usepackage{titlesec}
\usepackage[absolute,overlay]{textpos}

\begin{document}

\title{Self-Interaction of Super-Resonant Dark Matter}

\author{Shao-Song Tang}
\email{107552300734@stu.edu.cn}
\affiliation{School of Physical Science and Technology, Xinjiang University,Urumqi 830046, China}

\author{Murat Abdughani}
\email{mulati@xju.edu.cn}
\affiliation{School of Physical Science and Technology, Xinjiang University,Urumqi 830046, China}

\begin{abstract}
The $\Lambda$CDM model, while successful on large cosmological scales, faces challenges on small scales. A promising solution posits that dark matter (DM) exhibits strong self-interaction, enhanced through the narrow resonance or Sommerfeld effects. We demonstrate that the ``super-resonance" phenomenon, combining these effects, significantly amplifies the DM self-scattering cross section, enabling strong self-interactions for DM candidates in the $\mathcal{O}(100)$ GeV mass range. This mechanism also enhances the DM annihilation cross section, causing early kinetic decoupling that renders the standard Boltzmann equation inadequate. By implementing coupled Boltzmann equations, we achieve precise calculations of the relic density for super-resonant DM, aligning with observational constraints.
\end{abstract}

\date{\today}
\maketitle
%\tableofcontents
\newpage
\maketitle

\section{Introduction}\label{sec:Introduction}
Dark matter (DM) plays a large role in the formation of the structure of the universe, but the nature of DM remains undetermined. The $\Lambda$CDM model is one of the most popular models, which is very successful in explaining many currently observed phenomena in cosmology~\cite{Bahcall:1999xn,Planck:2018vyg,Riess:2019cxk,sDss:2005xqv,Springel:2005nw,Weinberg:2013agg,WMAP:2010qai}, such as Cosmic Microwave Background Radiation (CMB)~\cite{WMAP:2008lyn}, the Baryon Acoustic Oscillations (BAO)~\cite{Ding:2019hmw}, and Large Scale Structure of the Universe (LSS)~\cite{Blumenthal:1984bp, 10.1007/s11467-016-0583-4}. In particular, this theory of gathering ordinary matter through DM halos to produce the many galaxies and clusters of galaxies that exist a striking consistency with observation at present. 

Although the $\Lambda$CDM model has achieved remarkable success on large cosmological scales, it confronts several persistent challenges~\cite{Weinberg:2013aya, Bullock:2017xww,Sales:2022ich}, particularly the unresolved tensions between its predictions and observational data on small scales. A prominent manifestation emerges in the ``Core-Cusp" problem~\cite{deBlok:2009sp, Navarro:1995iw, Moore:1999gc} -- observations reveal DM density profiles resembling ``Core" distributions~\cite{Flores:1994gz}, contrasting with the ``Cusp"-like profiles predicted by collisionless CDM simulations. Another critical issue arises in the ``Too-Big-to-Fail" problem~\cite{Garrison-Kimmel:2014vqa, Papastergis:2014aba, Brook:2014hda}, where discrepancies persist between the simulated mass function of DM halos in small-scale cosmology and the observed mass distribution of galactic systems.

For the small scale challenges of $\Lambda$CDM, a potential resolution involves postulating self-interacting DM~\cite{Spergel:1999mh, Vogelsberger:2012sa, Kaplinghat:2015aga, Zhu:2021pad}, where the self-interaction manifests large scattering cross sections $\sigma_{\mathrm{SI}}/m_{\chi} \sim 10^{-1} \text{--} 10^{1}~\mathrm{cm}^2/\mathrm{g}$ on subgalactic scales~\cite{Colin:2002nk, Meneghetti:2000gm, Yang:2022hkm}. Both narrow resonance and Sommerfeld effects can significantly amplify DM self-scattering cross sections~\cite{Chu:2018fzy,Ibe:2008ye, Belanger:2024bro, Duch:2017nbe, Feng:2010zp, Arkani-Hamed:2008hhe}. The synergistic super-resonance phenomenon, incorporating both mechanisms, elevates these cross sections by several orders of magnitude (up to $10^{2}\text{--}10^{4}$), thereby achieving the required enhancement~\cite{Beneke:2022rjv, Csaki:2022xmu}. 

Simultaneously, within the super-resonance framework, the coexistence of resonance effects and Sommerfeld enhancements precludes gauge boson exchange between DM particles and intermediate resonance states under higher-order corrections of narrow-width expansion, nor does it permit exchange interactions between initial DM particles and final-state products~\cite{Beneke:2022rjv}. Consequently, the resonance and Sommerfeld effects demonstrate complete factorizability, meaning the super-resonance enhanced interaction cross section can be expressed as the product of resonance and Sommerfeld factors. This decomposition significantly facilitates the derivation of both annihilation and self-scattering cross sections for SRDM.

It should be emphasized that the super-resonance mechanism not only significantly enhances the DM annihilation cross section but also induces early kinetic decoupling of DM. This intertwining of kinetic and chemical decoupling processes invalidates the foundational assumption of kinetic equilibrium between DM particles and thermal bath particles in the standard Boltzmann equation~\cite{Edsjo:1997bg}. Consequently, the standard Boltzmann equation that only considers the first moment of the DM phase space distribution (number density) loses validity. This necessitates the implementation of coupled Boltzmann equations (cBEs)~\cite{vandenAarssen:2012ag, Duch:2017nbe, Kamada:2017gfc, Abe:2020obo} which incorporate the second moment (velocity dispersion) of the phase space density, or alternatively the adoption of full Boltzmann equation~\cite{Binder:2017rgn, Ala-Mattinen:2019mpa, Binder:2021bmg} to obtain refined results~\cite{Abdughani:2025rsw}. 

Moreover, the velocity-dependent enhancement of annihilation rates via super-resonance could elevate the DM annihilation cross section into Standard Model (SM) particles beyond the CMB constraints~\cite{Planck:2018vyg}. For example, Ref.~\cite{Bringmann:2016din} systematically demonstrates that, in light-mediator SIDM models, the Sommerfeld-enhanced (s)-wave DM annihilation rate is in tension with constraints from the CMB and indirect detection, thereby completely excluding the region of parameter space that yields the astrophysically relevant self-interaction cross section ($\langle\sigma_T\rangle_{30}/m_\chi \sim 0.1\text{--}10~\mathrm{cm}^2\mathrm{g}^{-1}$) required to address small-scale structure problems. To avoid this conflict, two potential solutions exist: (i) DM annihilate only to neutrino final state; (ii) the DM models operate within dark sector. In Ref.~\cite{Abdughani:2025rsw}, the former case have been studied and found that lower limit to the DM mass is at the order of $\mathcal{O} (100)$ GeV, which is too heavy to produce enough self-interaction cross section. Therefore, in this work, we focus on scenario (ii), studying the self-interaction of super-resonant DM (SRDM) model within a dark sector, and constrain its parameters with current experimental data.

The structure of this paper is organized as follows: In Sec.~\ref{sec:Boltzmann}, present the cBEs in dark sector. In Sec.~\ref{sec:Annihilation}, analyze the super-resonance enhanced annihilation cross sections through a benchmark SRDM model case study, providing accurate calculations of the relic density. In Sec.~\ref{sec:Self-interaction}, we describe the self-interacting properties of SRDM and summarize in Sec.~\ref{sec:Conclusion}. In Appendix~\ref{sec:app}, we provide detailed derivations of the annihilation and self-scattering cross sections for SRDM.

\section{Boltzmann equation beyond kinetic equilibrium}\label{sec:Boltzmann}

For thermal relic DM, during the early Universe epoch characterized by high cosmological temperatures, DM particles remained in dynamical equilibrium with other particles through continuous annihilation and scattering processes, forming a thermal bath~\cite{Beneke:2016ync, Harigaya:2016nlg}. As the Universe expanded and cooled, DM particles gradually decoupled from equilibrium and became cosmologically ``freeze-out"~\cite{Binder:2021bmg}, ultimately constituting approximately 26\% of the Universe's present total energy content~\cite{Bertone:2016nfn}. The decoupling process of thermal relic DM is systematically addressed through a standardized methodology -- precise calculation of the DM relic density by solving the Boltzmann equation~\cite{Edsjo:1997bg, Oncala:2021swy, Binder:2021bmg, Abe:2020obo}.

In most scenarios, solving the Boltzmann equation for DM relic density requires only consideration of the first hierarchy of the phase-space distribution, which necessitates the assumption of maintained kinetic equilibrium prior to chemical decoupling~\cite{Bringmann:2009vf}. Although SRDM belongs to the thermal relic category and generates its present abundance through thermal decoupling, the super-resonance mechanism alters both annihilation and self-interaction rates, thereby modifying the thermal decoupling dynamics. This modification induces premature kinetic decoupling that becomes entangled with chemical decoupling. Under such circumstances, calculations relying solely on the first hierarchy of the DM phase-space distribution lose accuracy. Consequently, incorporating the second hierarchy (i.e., DM velocity dispersion) or solving the complete Boltzmann equation becomes essential for achieving precise results~\cite{Duch:2017nbe, Kamada:2017gfc, Abe:2020obo, Liu:2023kat}.

When investigating the cosmological temperature evolution of DM particles, analysis is focused on their collective statistical behavior rather than individual particle dynamics. The kinetic evolution of DM during decoupling can be characterized by its phase-space density distribution function $f_\chi(\mathbf{x},\mathbf{p},t)$. By definition, this distribution function must satisfy:
\begin{equation}
    \frac{d f_{\chi}}{dt} = C[f_{\chi}] \, ,
\end{equation}
where $C[f_{\chi}]$ denotes the collision term.

Within the framework of Friedmann-Robertson-Walker (FRW) cosmology -- which posits a homogeneous and isotropic Universe -- the decoupling process of DM is rigorously described by the full Boltzmann equation (fBE): 
\begin{equation}
    E (\partial t - H p \, \partial p) f_{\chi} = C_{el}[f_{\chi}] + C_{ann}[f_{\chi}] \, ,
    \label{eq:FRW}
\end{equation}
where $H \equiv \dot{a}/a$ denotes the Hubble constant with $a$ being the scale factor, while $C_{\text{el}}$ and $C_{\text{ann}}$ represent the elastic scattering term and binary annihilation term respectively.

The treatment of the annihilation term is straightforward: neglecting Bose enhancement and Pauli blocking factors, it can be readily converted into terms involving the DM annihilation cross section, which we shall not elaborate further here. In contrast, the scattering term generally poses greater computational challenges. However, when chemical decoupling occurs in the highly non-relativistic regime of DM, the elastic scattering term undergoes significant simplification. 
When the typical momentum transfer per collision remains much smaller than the equilibrium average DM momentum, expanding $C_{\text{el}}$ to second order yields a simplified Fokker-Planck-type differential operator~\cite{Bertschinger:2006nq, Bringmann:2009vf, Binder:2016pnr}:
\begin{equation}
    C_{\text{el}} \simeq C_{\text{FP}} = \frac{E}{2} \gamma (T) \, [T E \partial^2_P + (2 T E / p + p + T p / E) \, \partial_p + 3] f_{\chi} \, .
    \label{eq:FP}
\end{equation}
where the momentum transfer rate $\gamma(T)$ is given by:
\begin{equation}
\gamma = \frac{1}{3 g_\chi m_{\chi} T} \int \frac{\mathrm{d}^3 k}{(2\pi)^3} g^{\pm}(\omega) \big[1 \mp g^{\pm}(\omega)\big] \int^{0}_{-4k_{\text{cm}}^2} \mathrm{d}t \, (-t) \frac{\mathrm{d}\sigma}{\mathrm{d}t} v\, .
\end{equation}
where $g_\chi$ denotes the internal degrees of freedom of the DM particle, $g^{\pm}(\omega) \equiv 1/[e^{\omega/T} \pm 1]$ encodes the quantum statistical distribution, and the differential cross section times relative velocity is defined as $\frac{\mathrm{d}\sigma}{\mathrm{d}t} v \equiv |\mathcal{M}|_{\chi f\leftrightarrow\chi f}^2/(64\pi k \omega m_\chi^2)$. The Mandelstam variable $k_{\text{cm}}^2$ follows the relativistic dispersion relation $k_{\text{cm}}^2 \equiv m_\chi^2 k^2/(m_\chi^2 + 2\omega m_\chi + m_f^2)$, with $k$ representing the total momentum of the DM particle.

Compared to the conventional number density Boltzmann equation (nBE) equation that considers only the first moment of the phase-space distribution (DM number density $n = g_\chi \int \mathrm{d}^3 p \, f_{\chi}/{(2\pi)^3}$), premature kinetic decoupling induces temperature decoupling where the DM particle temperature deviates from the background (dark sector or SM) thermal bath. Post kinetic decoupling, the DM temperature evolves as $T_\chi \propto a^{-2}$ while the background temperature follows $T \propto a^{-1}$. For SRDM, the velocity-dependent annihilation cross section will modify the chemical decoupling dynamics of DM. Consequently, the simplified nBE approach becomes inadequate for precise relic density calculations, necessitating the implementation of cBEs that incorporate the second moment of the phase-space distribution to faithfully track the DM evolution and determine the relic density~\cite{Abdughani:2025rsw}.

The comoving number density and velocity dispersion of DM are defined through the following relations:
\begin{equation}
\begin{cases}
\displaystyle
Y = \frac{n_{\chi}}{s}, \\[12pt]
\displaystyle
y \equiv \frac{m_{\chi}}{3 s^{2/3}} \left\langle \frac{p^2}{E} \right\rangle = \frac{m_{\chi}}{3 s^{2/3}} \frac{g_\chi}{n_\chi} \int \frac{\mathrm{d}^3 p}{(2\pi)^3} \frac{p^2}{E} f_{\chi},
\end{cases}
\end{equation}
where $s = (2\pi^2/45) g_{\text{eff}}^2 T^3$ represents the entropy density, and the DM temperature is defined through the relation $T_{\chi} \equiv y s^{2/3}/m_{\chi}$.

In the dark sector framework, DM doesn't couple directly to SM particles, but is in thermal equilibrium with the thermal background particle $(l)$ in the dark sector, i.e., $T_\chi = T_l$ before DM decoupling.  We define the temperature ratio as
\begin{equation}
    r = \frac{T_l}{T_{\gamma}},
\end{equation}
where $T_l$ is the temperature of the thermal background particles in the dark sector and $T_\gamma$ is the photon temperature in the SM sector. $T_\chi$ will ultimately be linked indirectly to $T_\gamma$ through the $\chi$-$l$ coupling as well as the temperature ratio $r$. Specifically, the temperature ratio $r$ affects the calculation of the relic abundance mainly through two aspects: (i) influencing the thermally averaged annihilation cross sections $\langle \sigma v \rangle_{T_\chi}$ and $\langle \sigma v \rangle_{2,T_\chi}$, which depend on $T_\chi$; (ii) affecting the scattering rate $\langle C_{\mathrm{el}} \rangle_2$ via the momentum transfer rate $\gamma(T_l)$, which directly impacts the kinetic decoupling of DM.

We consider that the inflaton couples differently to the visible and dark sectors, leading to them having different temperatures at the reheating epoch~\cite{Feng:2008mu}. Due to the extremely weak coupling between the two sectors, this ratio subsequently evolves only according to the relative entropy degrees of freedom in each sector. Although the temperatures of the two sectors evolve independently, the conservation of total entropy density ($d(sa^3)/dt = 0$) governs the evolution of their ratio~\cite{Zhu:2021pad,Berlin:2016gtr,Feng:2008mu}:
\begin{equation}
    r(T_\gamma) = r_{\rm RH} \left( \frac{g_s(T_\gamma)}{g_{s,\rm RH}} \right)^{1/3} \left( \frac{g_{s,\rm RH}^h}{g_{s}^h(T_l)} \right)^{1/3},
    \label{eq:entropy-conservation}
\end{equation}
where $r_{\rm RH}$ denotes the temperature ratio during the reheating era, and $g_{s,\rm RH}$ and $g_{s,\rm RH}^h$ represent the entropy degrees of freedom of the two sectors at that time, respectively. By tracking the evolution of the entropy degrees of freedom in the two sectors, one can determine the temperature ratio $r(T_\gamma)$ at any epoch.

Relativistic energy density in the dark sector affects the measurement of the effective number of relativistic neutrino species $N_{\rm eff}$, and contribution from $l$ can be expressed as~\cite{Zhu:2021pad}
\begin{equation}
    \Delta N_{\rm eff} = \frac{4}{7} \left( \frac{11}{4} \right)^{4/3} g_s^h r^4 .
    \label{eq:N_eff}
\end{equation}
For our model, with $g_{s,\rm BBN}^h = 1.75$, choosing $r_{\mathrm{RH}} \lesssim 0.58$ yields $r(T_\gamma^{\mathrm{BBN}}) \lesssim 0.5$, thereby ensuring that the impact on $N_{\mathrm{eff}}$ remains below the limit of $\Delta N_{\rm eff} \lesssim 0.26$~\cite{Planck:2018vyg} from BBN.
%(In our model, $r(T_\gamma^{\mathrm{BBN}}) > r(T_\gamma^{\mathrm{CMB}})$, therefore, if the BBN constraint is satisfied, the current CMB constraint is also.)

Furthermore, the energy density in the dark sector would also modify the Hubble expansion rate. Specifically, the modified Friedmann equation reads:
\begin{equation}
    H^2 = \frac{8 \pi G}{3} \left[ \rho_\gamma + \rho_\nu + \rho_{\rm baryon} + \rho_{\text{dark}} + \rho_{\Lambda} \right], \label{eq:hubble}
\end{equation}
where $\rho_{\text{dark}}$ denotes the total energy density in the dark sector, given by:
\begin{equation}
    \rho_{\text{dark}} = 2m_{\chi} s (Y + Y_{l}).
\end{equation}
In order to track the evolution of the injected energy density in the dark sector, we need to additionally consider the evolution equation $Y_l \equiv n_l / s = \rho_l / (m_\chi \, s)$~\cite{Binder:2017lkj}.

Integrating Boltzmann Eq.~\eqref{eq:FRW} over $g_\chi \int \, d^3p / (2 \pi)^3$ and $g_\chi \int p^2 / E^2 \, d^3p / (2 \pi)^3$, a set of cBEs for DM abundance tracking is~\cite{Binder:2017lkj}:
\begin{align}
    \frac{Y'}{Y} &= - \frac{s Y}{x \tilde H} {\langle \sigma v \rangle}_{T_{\chi}}, 
    \nonumber
    \\
    \frac{y'}{y} &= \frac{s Y}{x \tilde H} \left[ {\langle \sigma v \rangle}_{T_{\chi}} - {\langle \sigma v \rangle}_{2, T_{\chi}} \right] + \frac{1}{x \tilde H} \langle C_{\text{el}} \rangle_2 \ .
    \label{eq:cBE}
\end{align}
Meanwhile, dark radiation energy density can be obtained from
\begin{equation}
    \frac{Y_l'}{Y} = \frac{s Y}{x \tilde H} {\langle \sigma v \rangle}_{T_{\chi}} - \frac{H}{x \tilde H} \frac{Y_l}{Y} \ .
\end{equation}
Here $\tilde H$ represents a modified Hubble parameter:
\begin{equation}
    \tilde H \equiv \frac{H}{1 + 3 \tilde g(x)}, \quad \text{with} \quad \tilde g(x) = \frac{1}{3} \frac{T}{g^s_{\text{eff}}} \frac{\mathrm{d} g^s_{\text{eff}}}{\mathrm{d}T},
\end{equation}
where $g^s_{\text{eff}}$ denotes the effective entropy degrees of freedom. ${\langle \sigma v \rangle}_{T_{\chi}}$ is the thermally averaged annihilation cross section and ${\langle \sigma v \rangle}_{2,\, T_{\chi}}$ corresponds to a temperature-weighted variant of the thermally averaged annihilation cross section:
\begin{align}
    {\langle \sigma v \rangle}_{T_{\chi}}& \equiv \frac{g^2_{\chi}}{n^2_{\text{eq}}} \int \frac{\mathrm{d}^3 p \, \mathrm{d}^3 \tilde p}{(2\pi)^6} \sigma_{\text{anni}} v f_{\chi,\text{eq}}(p) f_{\chi,\text{eq}}(\tilde p),
    \\
    {\langle \sigma v \rangle}_{2, T_{\chi}}& \equiv \frac{g^2_{\chi}}{T_{\chi} \, n^2_{\text{eq}}} \int \frac{\mathrm{d}^3 p \, \mathrm{d}^3 \tilde p}{(2\pi)^6} \cdot \frac{p^2}{3E} \, \sigma_{\text{anni}} v f_{\chi,\text{eq}}(p) f_{\chi,\text{eq}}(\tilde p) \ .
\end{align}
The integration of elastic scattering term, $\langle C_{el} \rangle_2$, is defined as:
\begin{equation}
    \langle C_{el} \rangle_2 \equiv \frac{g_\chi}{3 n T_\chi} \int \frac{d^3p}{(2\pi)^3} \frac{p^2}{E^2} C_{el} \, .
\end{equation}
For highly non-relativistic DM, the Fokker-Planck approximation in Eq.~\eqref{eq:FP} enables the reduction $\langle C_{\text{el}} \rangle_2 \rightarrow 2\gamma(T)[y_{\text{eq}}/y - 1]$. Substituting this approximation into Eq.~\eqref{eq:cBE} yields the final cBEs capable of precisely calculating the relic density of DM. In summary, given the elastic scattering amplitude $\mathcal{M}_{\chi f \leftrightarrow \chi f}$ other than annihilation cross section $\sigma v$, more accurate predictions of the decoupling process and DM relic density are achieved.

\section{Super-resonant enhancement of the DM annihilation cross section}\label{sec:Annihilation}

The super-resonance mechanism exerts significant influence on DM annihilation, manifesting through two distinctive features: (i) the narrow resonance peak enhancement near resonance poles and (ii) the velocity Sommerfeld amplification at low velocities. These combined effects amplify the DM annihilation cross section by several orders of magnitude~\cite{Beneke:2022rjv}, thereby may modify the thermal decoupling dynamics of DM.

To quantify the super-resonance effects on DM annihilation, we consider a simple SRDM model in a dark sector comprising: a Dirac fermion DM candidate $\chi$, real vector bosons $A$ and $V$ with masses $m_A \ll m_\chi$ and $m_V \simeq 2m_\chi$ respectively, and the massless spin-half thermal background particle $l$. The vector boson $V$ operates as the intermediate resonance state, whereas the light vector boson $A$ mediates the Sommerfeld enhancement through long-range interactions. The minimal interaction Lagrangian can be written as (SM neutrino final state have been studied in Ref.~\cite{Abdughani:2025rsw}):
\begin{equation}
    \mathcal{L} \supset - g_A \overline{\chi} \gamma^{\mu} \chi A_{\mu} - g_V \overline{\chi} \gamma^{\mu} \chi V_{\mu} - g_l \overline{l} \gamma^{\mu} l V_{\mu} \, ,
\end{equation}
where $g_A$, $g_V$, and $g_l$ are coupling constants. This model is phenomenologically viable, and a gauge-invariant extension can be constructed by introducing additional fields. In this SRDM model, DM particles annihilate through two distinct pathways: (i) via $s$-channel exchange of the resonant vector boson $V$ annihilate into $l$ final states, (ii) directly annihilate into gauge bosons $A$.

The annihilation rate for the process $\chi \overline{\chi} \to A \overline{A}$ is given by:
\begin{equation}
    \sigma v_{\text{rel}} (\chi \overline{\chi} \to A \overline{A}) = \frac{4 \pi \alpha^2_A}{m^2_{\chi}} \times S_{\text{SF}}(v_{\text{rel}}),
\end{equation}
where $\alpha_A \equiv g^2_A/(4\pi)$ denotes the effective coupling constant, and $v_{\text{rel}} = 2v$ represents the relative velocity between annihilating particles. The velocity-dependent Sommerfeld enhancement factor $S_{\text{SF}}$ for $s$-wave annihilation reads~\cite{Cassel:2009wt}:
\begin{equation}
    S_{\text{SF}}(v) = \frac{\pi \alpha_A \sinh\left(\dfrac{12 m_{\chi} v}{\pi m_A}\right)}{v \left[ \cosh\left(\dfrac{12 m_{\chi} v}{\pi m_A}\right) - \cos\left(2\pi \sqrt{\dfrac{6 m_{\chi} \alpha_A}{\pi^2 m_A} - \dfrac{36 m^2_\chi v^2}{\pi^4 m^2_A}} \right) \right]}.
    \label{eq:sommerfeldFactor}
\end{equation}

For the annihilation channel $\chi \overline{\chi} \to V \to l \overline{l}$, the super-resonance mechanism allows complete factorizability between resonance and Sommerfeld effects~\cite{Beneke:2022rjv}, leading to the cross section expression~\cite{Abdughani:2025rsw}~(see Appendix~\ref{sec:app} for detailed derivation):
\begin{equation}
    \sigma v_{\text{rel}} (\overline{\chi} \chi \to \overline{l} l) = \frac{g^2_V}{4m^2_{\chi}}\frac{m_{\chi}\Gamma_{V,\bar{l}l}/2}{(\frac{1}{4}m_{\chi}v^2 - \delta M)^2 + \Gamma^2_V/4} \times S_{\text{SF}}(v)\,,
    \label{eq:superResonance}
\end{equation}
where $\delta M \equiv m_V - 2m_{\chi}$. The resonance peaks at a certain velocity $v=v_{res}$, i.e. when $\delta M=\frac{1}{4}m_{\chi}v_{res}^2$. $\Gamma_{V,ll}$ represents partial decay width of $V$ to the $l\overline{l}$ final states, and it is
\begin{equation}
    \Gamma_{V,\bar{l} l} = \frac{g_l^2}{12 \pi} m_V \ .
    %\left( 1 + \frac{2 m_l^2}{m_V^2} \right) \sqrt{1 - \frac{4 m_l^2}{m_V^2}}.
    \label{eq:Gamma_ll}
\end{equation}
$\Gamma_V$ denotes the total decay width of $V$, and it decomposes as
\begin{equation}
    \Gamma_V = \Gamma_{V,\bar{l}l} + \Gamma_{V,\bar{\chi}\chi} + \Gamma_{V,AA},
\end{equation}
where $\Gamma_{V,\bar{\chi}\chi}$ is the decay width of $V$ to DM pair if kinematically allowed, and it is 
\begin{equation}
    \Gamma_{V,\bar{\chi} \chi} = \frac{g_V^2}{12 \pi} m_V \left( 1 + \frac{2 m_\chi^2}{m_V^2} \right) \sqrt{1 - \frac{4 m_\chi^2}{m_V^2}} = \frac{g_V^2}{8\pi}\sqrt{2m_V \delta M} \ .
    \label{eq:Gamma_xx}
\end{equation}
Light mediator $A$ does not couple to $V$ at the tree level, thus the decay width of $V$ to $A$ pair, $\Gamma_{V,AA}$, is loop suppressed. To match the observed relic density and achieve a large self-interaction cross section, $g_l$ must be significantly smaller than $g_V$. Thus the approximation
\begin{equation}
    \Gamma_{V} \simeq \Gamma_{V,\bar{\chi} \chi}
    \label{eq:Gamma}
\end{equation}
is plausible for convenience.

The real vector $A$ decays to dark radiation through a DM loop, and the decay width is approximately $\Gamma_A \simeq 10^{-5} m_\chi \alpha_A^2 g_V^2 g_l^2$. Ensuring that $A$ decays before BBN requires $\Gamma_A > 10^{-25}$ GeV . 

Furthermore, precise solution of the cBEs requires additional specification of the scattering amplitudes for $\chi l \leftrightarrow \chi l$ process, which can be expressed as
\begin{equation}
    {|\mathcal{M}|}^2_{\chi \, l \leftrightarrow \chi \, l} = \frac{16{(g_l g_V)}^2 {(2 m_\chi / m_A)}^2}{(2 m_\chi / m_A) t - 4 m^2_\chi} \beta_1 (\omega , t) \, ,
\end{equation} 
where,
\begin{equation}
    \beta_1 (\omega , t) = 8 m^2_\chi \omega^2 + 4 m^2_\chi (\omega / m_\chi + 1/2) \, t + t^2 \, ,
\end{equation}
Here, the squared amplitudes ${|\mathcal{M}|}^2_{\chi l \leftrightarrow \chi l}$ is obtained by summing over all initial and final spin states of particles, including both particle and antiparticle.

The SRDM model in Eq.~\eqref{eq:superResonance} have three mass and three coupling free parameters. To facilitate the parameter scan, the value of $m_A$ is constrained by the Sommerfeld resonance condition
\begin{equation}
    m_A = \frac{6 \alpha_A m_{\chi}}{n^2 \pi^2},\quad\quad \text{with} \quad n=1,2,3...
    \label{eq:mA}
\end{equation}
Actually, we firstly undergo the parameter scan of self-interaction of SRDM in Sec.~\ref{sec:Self-interaction}, where $g_l$ in not involved. And then, at the range of $3\sigma$ region constrained by observational data, numerical calculations of Boltzmann equations are carried out by appropriately modifying the Mathematica package DRAKE~\cite{Binder:2021bmg}.

\begin{figure}
    \centering
    \includegraphics[width=0.98\textwidth]{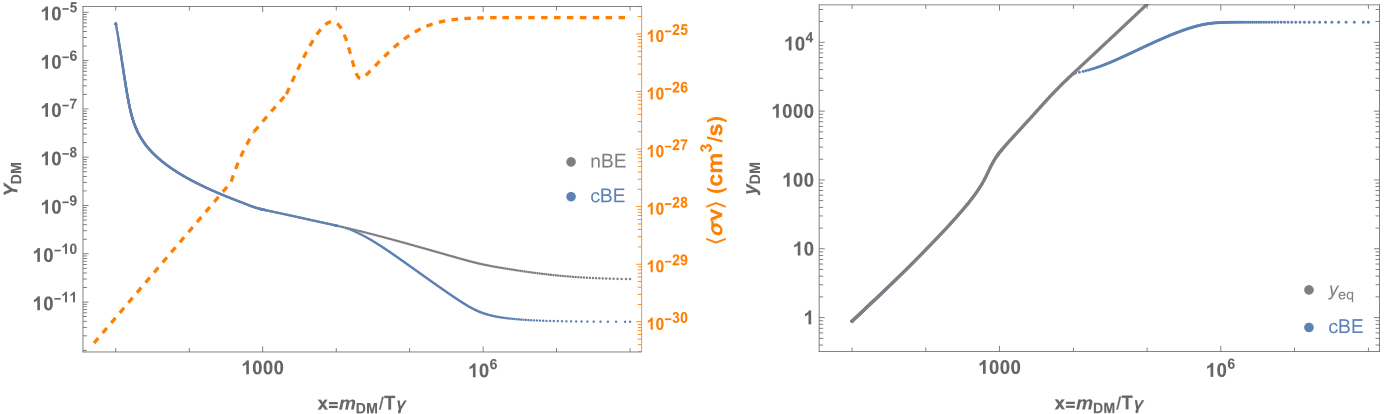}
    \caption{Thermal evolution of the $Y$ (left vertical axis of left panel), $\langle \sigma v \rangle$ (right vertical axis of left panel), and effective temperature $y$ (right panel) as a function of $x=m_\chi/T_\gamma$. Blue and gray dotted lines for the evolution of $Y$ and $y$ are obtained with solving the cBEs and the nBE, respectively.}
    \label{fig:relic}
\end{figure}

In TABLE~\ref{tab:Benchmark}, we show the parameters of a best-fit point (red stars in the Fig.~\ref{fig:result}) with which correct relic density of $\Omega_{\chi} h^2_{\text{cBE}} = 0.118829$ is obtained from cBEs, while value obtained from nBE is $\Omega_{\chi} h^2_{\text{nBE}} = 0.908778$. In FIG.~\ref{fig:relic}, thermal evolution of the $Y$ (left vertical axis of left panel), $\langle \sigma v \rangle$ (right vertical axis of left panel), and effective temperature $y$ (right panel) as a function of $x=m_\chi/T_\gamma$. Blue and gray dotted lines for the evolution of $Y$ and $y$ are obtained with solving the cBEs and the nBE, respectively.

%\tss{For the benchmark parameters listed in Table~\ref{tab:Benchmark}, kinetic decoupling occurs at $x = 10^4$ (right panel, where $y_{\text{DM}}$ departs from $y_{\text{eq}}$), leading to a temperature evolution described by $T_\chi = T_l^2 / T_l^{\text{kd}}$. This behavior contradicts the basic assumption of nBEs, rendering the nBEs inaccurate and ultimately resulting in a relic density ratio of $\Omega_{\chi} h^2_{\text{nBE}} / \Omega_{\chi} h^2_{\text{cBE}} \simeq 7.65$. At low velocities, the Sommerfeld effect enhances the DM annihilation cross section. For $x > 2.2 \times 10^4$, $S_{\text{SF}}(v)$ becomes significant, further increasing $\langle \sigma_{\text{anni}} v \rangle$ and triggering a period of reannihilation. This process reduces the comoving abundance of $\chi$ by more than an order of magnitude until final freeze-out is reached around $x \sim 2 \times 10^6$. The resulting relic density, $\Omega_{\chi} h^2_{\mathrm{cBE}} = 0.118829$, is in excellent agreement with the observed DM abundance.}

Both from the annihilation cross section in the left panel and evolution of $y_{\rm DM}$ in the right panel, it can be seen that resonance factor begin to take effect from $ x \simeq 10^3$, and lead to the kinetic decoupling ($y_{\rm DM}$ departs from $y_{\rm eq}$) at resonance peak of $x \simeq 10^4$. At low velocities, the Sommerfeld effect enhances the DM annihilation cross section. After $x \simeq 2 \times 10^4$, $S_{\text{SF}}(v)$ becomes significant, further increasing $\langle \sigma v \rangle$ and triggering a period of reannihilation. In return, obtained relic density from cBEs is about one order smaller than that of nBE.

\begin{table}[t]
    \centering
    \begin{tabular}{ccccccccc}
\hline\hline
Parameters & $m_\chi$  & $v_{res}$ & $m_A$ & $\alpha_A$ & $g_V$ & $g_l$ & $\Omega_{\chi} h^2_{\text{cBE}}$ & $\Omega_{\chi} h^2_{\text{nBE}}$\\
\hline
Values & 110 GeV & \quad 0.0065 &\quad $\simeq6 \alpha_A m_\chi/\pi^2$ &\quad 1/4$\pi$ &\quad 0.031 &\quad $6.93\times10^{-7}$ &0.118829&0.908778 \\
\hline \hline
\end{tabular}
\caption{Benchmark parameters for the SRDM model and corresponding relic densities results from cBEs and nBE, with the best-fit point marked by red stars in Fig.~\ref{fig:result}.}
\label{tab:Benchmark}
\end{table}

In addition, injected dark radiation $\rho_{\rm dark}$ in Eq.~\eqref{eq:hubble} may influence the Hubble expansion rate compared to the standard $\Lambda{\rm CDM}$ cosmology. However, as shown in Fig.~\ref{fig:relic}, DM annihilation stop before the matter-radiation equality, and thus have no deviation from current CMB data obtained Hubble constant~\cite{Binder:2017lkj,Planck:2018vyg}.

\section{Self-interaction of the SRDM}\label{sec:Self-interaction}

Simulations of collisionless CDM and SIDM demonstrate that DM self-interactions can resolve the mass deficit of DM halos on small scales. A self-scattering cross section $\sigma_{\text{SI}}/m_\chi \gtrsim 0.5~\text{cm}^2/\text{g}$ addresses the too-big-to-fail problem in Milky Way satellites and Local Group dwarfs~\cite{Zavala:2012us, Elbert:2014bma}, while cored profiles in galaxy clusters require $\sigma_{\text{SI}}/m_\chi \sim 0.1~\text{cm}^2/\text{g}$~\cite{Tulin:2017ara, Elbert:2016dbb, Kaplinghat:2015aga}.

The self-scattering cross section in SIDM models exhibits strong dependence on specific DM particle physics scenarios. For conventional weakly interacting massive particles (WIMPs) -- where masses and couplings reside at the weak scale -- self-interactions remain negligible ($\sigma_{\text{SI}}/m_\chi \ll 0.1~\text{cm}^2/\text{g}$) even with self-interactions, falling orders of magnitude below values required. This prompted discussions of models involving Breit-Wigner resonance effects~\cite{Chu:2018fzy,Ibe:2008ye, Belanger:2024bro, Duch:2017nbe} and Sommerfeld enhancements~\cite{Feng:2010zp, Arkani-Hamed:2008hhe} for boosting DM self-interactions for specific velocity regions. But mass favored regions in these model still reside in MeV-GeV range.

The super-resonance phenomenon -- combining both mechanisms -- holds potential to alleviate small scale tensions for GeV-TeV scale massive DM candidates through amplifying both annihilation and self-interaction cross section, and exhibiting favorable velocity dependence. Since SRDM follows a Boltzmann distribution, self-interactions are naturally suppressed at high velocities, manifesting as negligible interaction strengths for relativistic DM particles. This ensures that SRDM only modifies CDM predictions on small scales while preserving CDM's successful large-scale structure formation -- crucial for retaining CDM's well-established cosmological predictions~\cite{Planck:2018vyg,Springel:2005nw}. The velocity-dependent suppression mechanism thus provides a natural cutoff, reconciling enhanced small-scale self-interactions with undisturbed large-scale $\Lambda$CDM behavior.

Following the self-scattering formalism for resonant self-interaction DM models in Ref.~\cite{Chu:2018fzy} and the super-resonance framework from Ref.~\cite{Beneke:2022rjv}, DM self-scattering viscosity cross section can be expressed as the product of the narrow resonance effect and the Sommerfeld effect~(self-scattering cross section also can be obtained as in Appendix~\ref{sec:app}):
\begin{equation}
   \sigma_{\text{SI}} = \frac{8\pi w}{m_\chi \delta M} \cdot \frac{\Gamma_{V,\bar{\chi}\chi}^2/4}{(\frac{1}{4}m_\chi v^2 - \delta M)^2 + \Gamma_{V}^2/4} \times S_{\text{SF}}^2(v),
    \label{eq:SI}
\end{equation}
where $w$ represents the symmetry factor:
\begin{equation}
    w = \frac{2 J_V + 1}{(2 J_\chi + 1)^2},
\end{equation}
where $J_V$ and $J_\chi$ denote the spins of $V$ and $\chi$ respectively. For our benchmark model with $J_V = 1$ and $J_\chi = 1/2$, the symmetry factor becomes $w = 3/4$. Full width $\Gamma_V$ can be replaced with partial width $\Gamma_{V, \bar \chi \chi}$ as depicted in Eq.~\eqref{eq:Gamma}.

It is conventionally assumed that DM follows a Maxwell-Boltzmann velocity distribution, with the maximum velocity cutoff set by the galactic escape velocity $v_{\text{max}}$. Particles exceeding this threshold cannot be gravitationally bound to the host galaxy. The thermally averaged scattering cross section $\langle \sigma v \rangle$ is then given by:
\begin{equation}
    \langle \sigma v \rangle = \int^{v_{\text{max}}}_0 f(v, v_0) \, \sigma v \, \mathrm{d}v, \quad \text{with} \quad f(v, v_0) = \frac{4v^2 e^{-v^2/v_0^2}}{\sqrt{\pi} v_0^3},
    \label{eq:Maxwell-Boltzmann}
\end{equation}
where $v_0$ relates to the galactic mean velocity $\langle v \rangle$ through $v_0 \simeq \sqrt{\pi} \langle v \rangle/2$. Notably, the escape velocity $v_{\text{max}}$ satisfies $v_{\text{max}} \gg \langle v \rangle$ in typical galactic potentials.

We perform uniform a parameter scan in the range of shown in TABLE~\ref{tab:prior}. $\alpha_A$ is taken discrete values for time-saving, and it has minor effect to the results. Total $\chi^2 $ is calculated with
\begin{equation}
    \chi^2_{\rm tot} = \chi^2_{\rm anni} + \chi^2_{\rm SI} \ ,
    \label{eq:chi2}
\end{equation}
$\chi^2$ calculated for deviation from correct relic density is
\begin{equation}
    \chi^2_{\rm anni} = \frac{(\Omega h^2_{\rm cBE}-0.118)^2}{0.002^2+(0.1 \times \Omega h^2_{\rm cBE})^2} \ ,
\end{equation}
where $\Omega h^2_{\rm cBE}$ is relic density obtained with cBEs. 0.118 and 0.002 are central value and statistical error from Planck observation~\cite{Planck:2018vyg}. We also assumed 10\% of theoretical uncertainty.

\begin{table}
    \centering
    \begin{tabular}{ccccccc}
    \hline \hline
        Parameters & $m_\chi$ & $v_{res}$ & $g_V$ & $g_l$ & n & $\alpha_A$ \\
        Range &\quad $10\sim1000$ GeV &\quad $0.001\sim0.01$ &\quad $0.01\sim0.1$ &\quad $10^{-7}\sim10^{-5}$ & 1,2 & $1/4\pi,10^{-2},10^{-3}$ \\
        Prior & flat & flat & flat & log & discrete & discrete \\ 
    \hline \hline
    \end{tabular}
    \caption{Ranges and priors for input parameters adopted in the scans.}
    \label{tab:prior}
\end{table}

\begin{figure}
    \centering
    \includegraphics[width=0.6\textwidth]{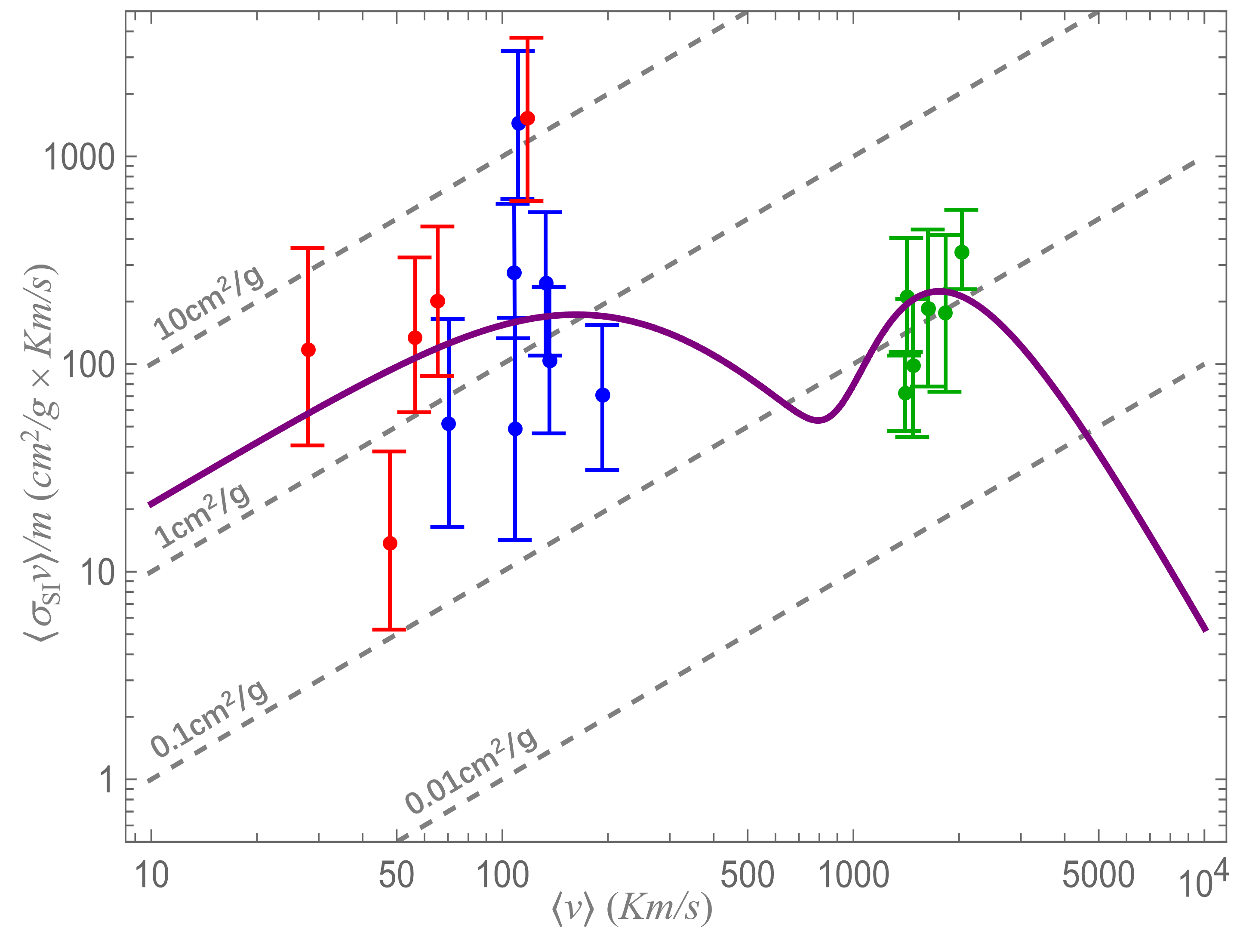}
    \caption{Velocity-averaged self-scattering cross section per unit mass $\langle \sigma_{\text{SI}}v \rangle/m_\chi$ as a function of galactic mean velocity. The purple solid curve is from our SRDM model with best-fit parameters. Gray dashed guidelines denote contours of constant $\langle \sigma_{\text{SI}}v \rangle/m_\chi$ values. Colored data points show converted $\langle \sigma_{\text{SI}}v \rangle/m_\chi$ constraints derived via semi-analytic methods from multi-galactic observations: five red points correspond to dwarf galaxies in the THINGS sample~\cite{Oh:2010ea}, six green points represent galaxy clusters from Newman et al.~\cite{Newman:2012nw, Newman:2012nv}, and seven blue points indicate low-surface-brightness (LSB) spiral galaxies in the Kuzio de Naray sample~\cite{KuziodeNaray:2007qi}.}
    \label{fig:SI}
\end{figure}

For self-interaction cross section, reduced $\chi^2$ is obtained with
\begin{equation}
\chi^2_{\rm SI} = \sum_{i=1}^{17}
\begin{cases}
    \frac{(\sigma^{\rm SI}_i - \sigma^0_i)^2}{(\sigma^{\rm +}_i)^2}\quad \text{if} \, \sigma^{\rm SI}_i \geq \sigma^0_i \ ,\\
    \frac{(\sigma^{\rm SI}_i - \sigma^0_i)^2}{(\sigma^{\rm -}_i)^2}\quad \text{if} \, \sigma^{\rm SI}_i < \sigma^0_i \ .
\end{cases}
\end{equation}
where $\sigma^{\rm SI}$ is $\langle \sigma_{\rm SI} v \rangle / m_\chi$ predicted from our model correspond to $i$-th data for certain velocity. $\sigma^0$ is central value of observational data, while $\sigma^{\rm +}$ and $\sigma^{\rm -}$ are the upper and lower error bounds of the observational data, respectively. Data for calculating $\chi^2_{\rm SI}$, shown in Fig.~\ref{fig:SI}, are obtained via semi-analytic methods from multi-galactic observations~\cite{KuziodeNaray:2007qi}: five red points correspond to dwarf galaxies in the THINGS sample~\cite{Oh:2010ea}, six green points represent galaxy clusters from Newman et al.~\cite{Newman:2012nw, Newman:2012nv}, and seven blue points indicate low-surface-brightness (LSB) spiral galaxies in the Kuzio de Naray sample~\cite{KuziodeNaray:2007qi}.

As stated in Sec.~\ref{sec:Annihilation}, we firstly undergo the parameter scan of self-interaction of SRDM where $g_l$ is not involved. And then, at the range of $3\sigma$ region constrained by observational data, numerical calculations of Boltzmann equations are carried out. After the calculation of $\chi^2_{\rm SI}$, $\chi^2_{\rm anni}$ can always be minimized by adjusting the parameter $g_l$.

\begin{figure}
    \centering
    \includegraphics[width=0.48\linewidth]{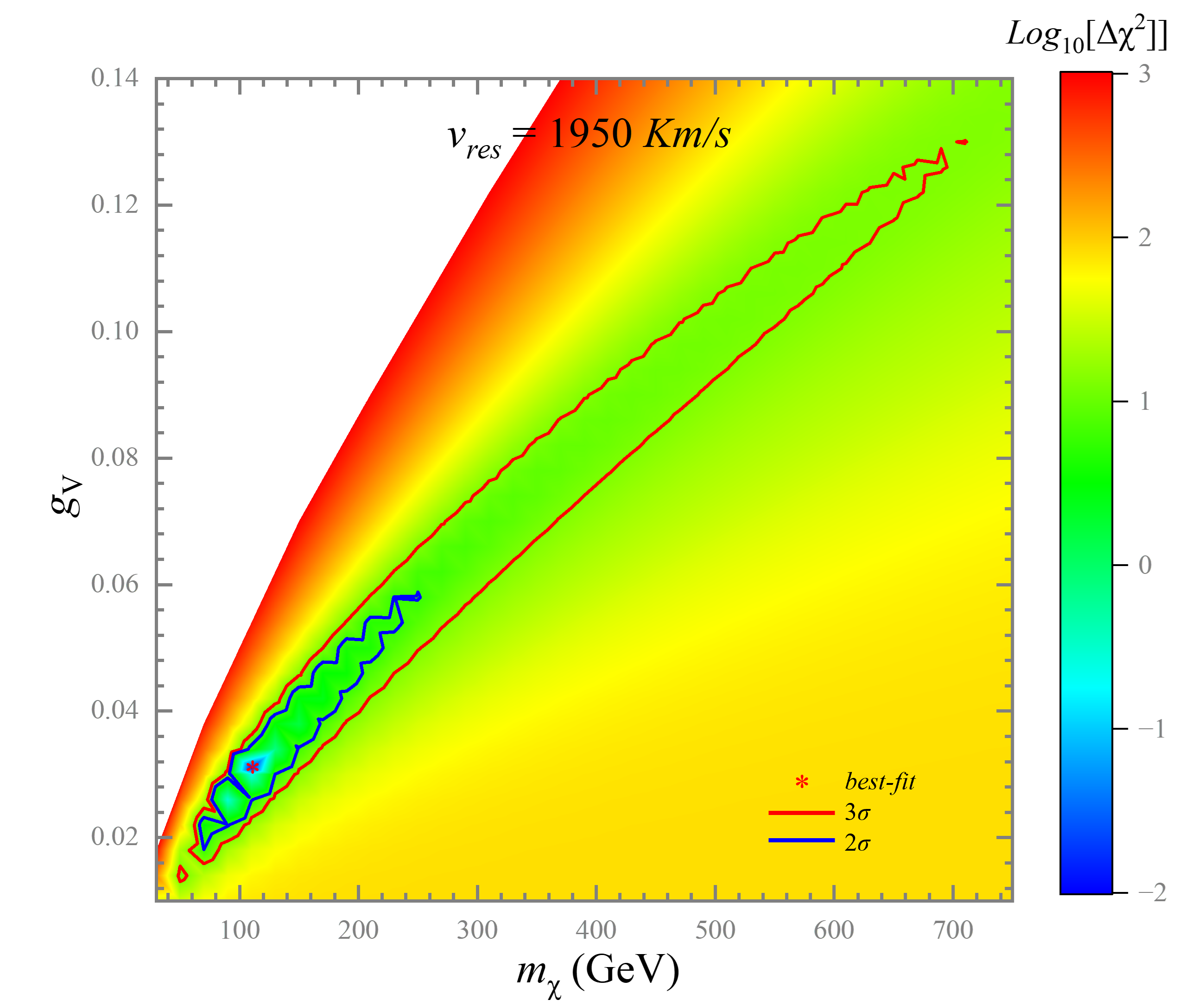}
    \includegraphics[width=0.48\linewidth]{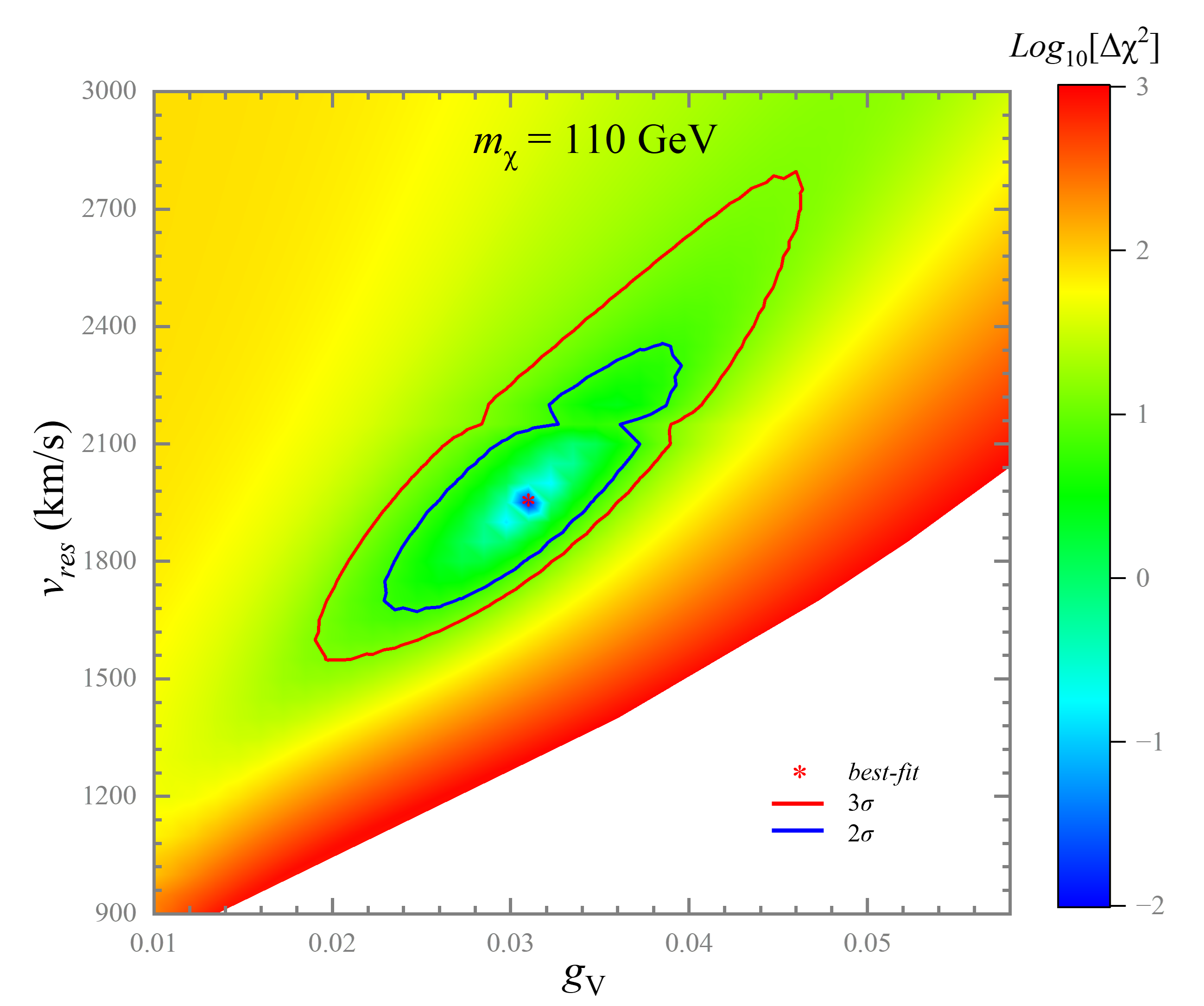}
    \includegraphics[width=0.48\linewidth]{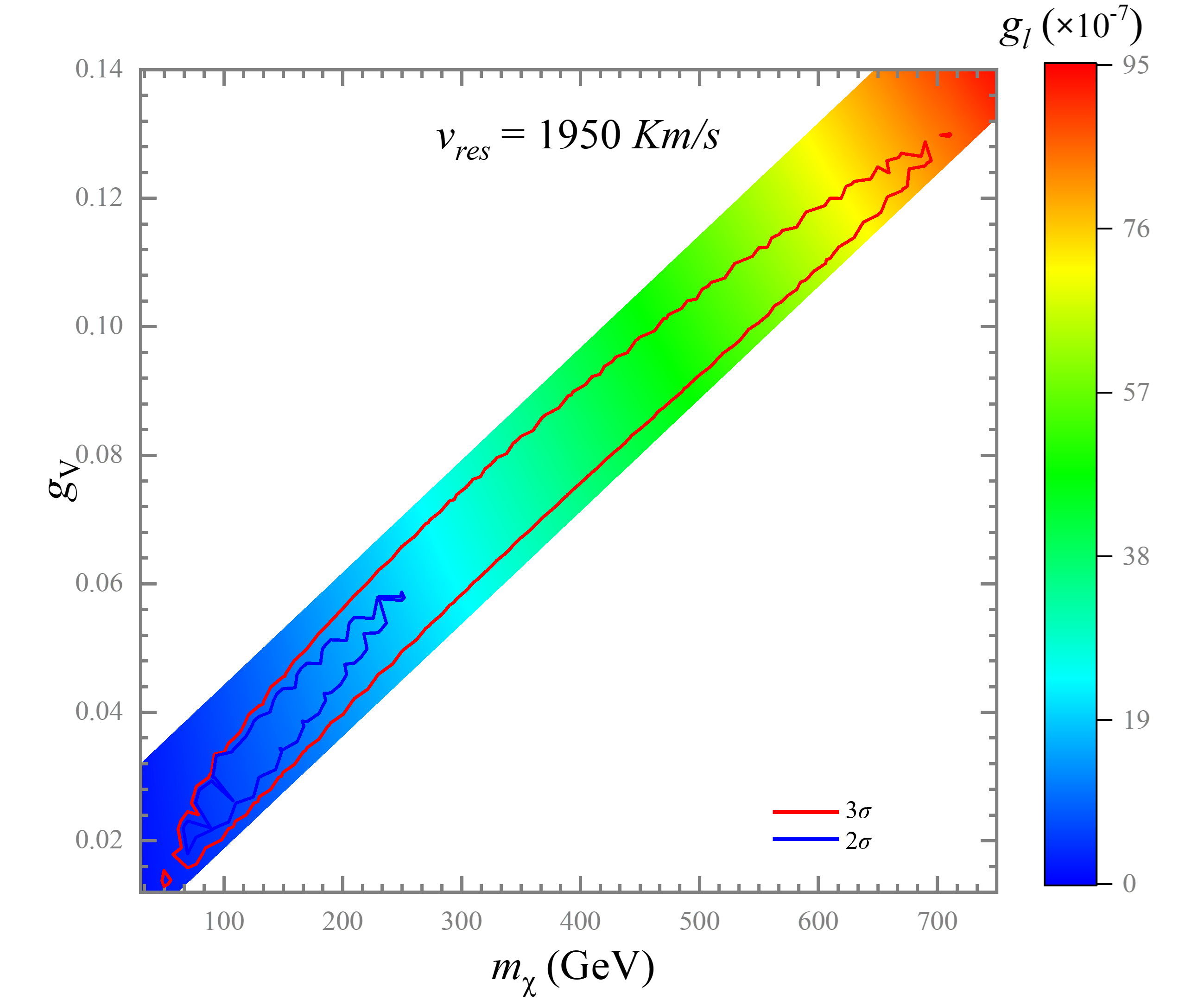}
    \caption{Upper panels show $\Delta \chi^2$ for two different parameter spaces, while $g_l$ as a color-bar shown in lower panel. Red stars are best-fit point with $\Delta \chi^2 = 0$ (assuming $\chi^2_{\rm anni} = 0$ for upper panels). Blue and red contours represent $2\sigma$ and $3\sigma$ limit with $\Delta \chi^2 < 6$ and $\Delta \chi^2 < 13$, respectively. Inserted texts $m_\chi = 110$~GeV and $v_{res} = 1950$~km/s are best-fit parameters.}
    \label{fig:result}
\end{figure}

For discrete values of $n$ and $\alpha_A$ given in Table~\ref{tab:Benchmark}, we have obtained minimum of $\chi^2_{\rm tot} = 17.2$ for $n = 1$ and $\alpha_A = 1/4\pi$. In Fig.~\ref{fig:result}, we show the corresponding scanned results in different parameter space. Red stars are best-fit point with $\Delta \chi^2 \equiv \chi^2_{\rm tot} - min(\chi^2_{\rm tot})= 0$ (assuming $\chi^2_{\rm anni} = 0$ for upper panels). Blue and red contours represent $2\sigma$ and $3\sigma$ limit with $\Delta \chi^2 < 6$ and $\Delta \chi^2 < 13$, respectively. Best-fit DM resonance velocity $v_{res}$ is 1950~km/s, and which comply with galaxy clusters' mean velocity (green data in Fig.~\ref{fig:SI}). DM mass is constrained to $\sim70-250$~GeV within $2\sigma$, while its range is in $\sim50-700$~GeV if within $3\sigma$ with best-fit mass of 110~GeV. 

Thus, in our SRDM model, self-interacting cross section can be greatly enhanced, and DM mass can be as heavy as $\mathcal{O} (100)$~GeV, while other works show at most DM mass of $\mathcal{O} (10)$~GeV to meet with observational data: in Ref.~\cite{Chu:2018fzy}, the RSIDM model yields an upper limit on the DM mass of approximately 40 GeV; The study in Ref.~\cite{Aboubrahim:2020lnr} investigates scenarios where DM annihilates into the dark sector via a dark photon mediator, also accounting for DM self-interactions. It suggests a preferred mass range between 1–10 GeV; Ref.~\cite{Chauhan:2017eck} considers a Majorana fermion DM model with diagonal interactions via a boson carrying dark $U(1)_D$ charge, showing that such a setup can simultaneously satisfy constraints from self-interaction and relic abundance, with a corresponding mass parameter space spanning from 10 keV to 1 MeV; Ref.~\cite{Wang:2023wbw} demonstrates that after Sommerfeld enhancement, self-interacting boosted DM can comply with small-scale structure constraints, and its mass is bounded above at about 10 GeV. Therefor, the model opens a region of SIDM parameter space at higher masses that was previously non-viable. This extension is a direct consequence of the super-resonant dynamics operating within the dark sector.

To quantitatively assess the advantages of the SRDM model, we compute the total goodness of fit, $\chi^{2}_{\mathrm{tot}}$, using the same methodology and benchmark parameter choices as in Ref.~\cite{Chu:2018fzy}, which features resonance-only enhancement, and Ref.~\cite{Kim:2021bmx}, which features Sommerfeld-only enhancement. For these two scenarios, we obtain $\chi^{2}_{\mathrm{tot}} = 20.3$ and $\chi^{2}_{\mathrm{tot}} = 26.4$, respectively. In contrast, the SRDM model yields a lower value, $\chi^{2}_{\mathrm{tot}} = 17.2$, indicating an improved overall fit to the combined astrophysical data. This improvement originates from the simultaneous presence of resonance and Sommerfeld effects in the dark matter self-interaction, which enhances the velocity dependence of the self-scattering cross section and enables a more accurate description of observations across multiple astrophysical scales.

As shown in Fig.~\ref{fig:SI}, purple solid line shows the velocity-averaged self-scattering cross section per unit mass of SRDM as a function of galactic mean velocity for best-fit point parameters. The resonant enhancement peaks at $\langle v\rangle = 1950~\text{km/s}$ (galaxy cluster scales), elevating $\langle\sigma_{\text{SI}}v\rangle/m_\chi$ to $\sim 0.1~\text{cm}^2/\text{g}$ -- consistent with cluster observations~\cite{Newman:2012nw}. For dwarf galaxies and LSB systems with lower velocities ($\langle v\rangle \lesssim 100~\text{km/s}$), the Sommerfeld-dominated regime amplifies $\sigma_{\text{SI}}/m_\chi$ to $1\text{--}3~\text{cm}^2/\text{g}$, matching their observational constraints~\cite{deBlok:2009sp,KuziodeNaray:2007qi}. This dual-scale agreement demonstrates the SRDM framework's ability to simultaneously resolve small-scale anomalies across different galactic systems.

Our systematic scan of the SRDM parameter space and comparison between best-fit parameters and multi-scale astrophysical observations demonstrate that heavy DM candidates in the GeV mass range can produce significantly enhanced self-scattering cross sections through the super-resonance mechanism, thereby effectively alleviating the small-scale structure problems of the $\Lambda$CDM model. However, under the current theoretical framework, limited by the upper bound on the light mediator coupling strength $\alpha_A$, the self-interaction cross section achieved via the super-resonance mechanism for TeV-scale DM remains well below the level required by astronomical observations. Consequently, for TeV-scale DM candidates, it is necessary to explore physical mechanisms beyond Sommerfeld enhancement and resonance effects---such as non-Abelian interaction structures~\cite{Tran:2023lzv,Abe:2020mph,Boddy:2014yra}, scalar mediator exchanges~\cite{Borah:2021rbx}, or composite DM models~\cite{Cline:2013zca}---to enhance the self-interaction strength, which would be an interesting research direction.

\section{Conclusion}\label{sec:Conclusion}
In this study, we have explored the self-interaction properties of SRDM, a framework that combines narrow resonance and Sommerfeld enhancement effects to significantly amplify the DM self-scattering and annihilation cross sections. This super-resonance mechanism provides a compelling solution to the small-scale challenges faced by the $\Lambda$CDM model by enabling strong self-interactions for DM candidates in the GeV mass range. Our analysis demonstrates that the SRDM model can enhance self-scattering cross sections to consistent with observational constraints from dwarf galaxies, low-surface-brightness spiral galaxies, and galaxy clusters.

The super-resonance effect also profoundly impacts the DM annihilation process, leading to early kinetic decoupling that intertwines with chemical decoupling. This necessitates the use of cBEs to accurately calculate the DM relic density, as the standard Boltzmann equation becomes inadequate.

Through a systematic parameter scan, we identified a best-fit point with a DM mass of $m_\chi = 110~\text{GeV}$, a resonance velocity of $v_{res} = 1950~\text{km/s}$, and coupling parameters that satisfy both relic density and self-interaction constraints within a $2\sigma$ range of $m_\chi \sim 70\text{--}250~\text{GeV}$. Standard nBE overestimates the relic density by nearly an order of magnitude compare to the value obtained with cBEs at the best-fit point.
Unlike previous models limited to DM masses of $\mathcal{O}(10)~\text{GeV}$, our SRDM framework extends the viable mass range to $\mathcal{O}(100)~\text{GeV}$, offering a robust mechanism to address small-scale structure anomalies while preserving the large-scale successes of $\Lambda$CDM cosmology.

\section*{Acknowledgements}
This work is supported by the Natural Science Foundation of Xinjiang Uygur Autonomous Region of China (No. 2025D01C48), Tianchi Talent Project of Xinjiang Uygur Autonomous Region of China, and the National Natural Science Foundation of the People’s Republic of China (No. 12303002).

\appendix

\section{Derivation of DM Annihilation and Self-Scattering Cross Sections}
\label{sec:app}

For the SRDM model considered in Sec.~\ref{sec:Annihilation}, the interaction between $\chi$ and the intermediate resonance state $V$ is given by:
\begin{equation}
    g_V \overline{\chi} \gamma^{\mu} \chi V_{\mu} + \text{h.c.}
\end{equation}

According to the optical theorem~\cite{Beneke:2003xh,Beneke:2004km}, the total DM annihilation cross section can be obtained from the forward scattering amplitude $\mathcal{M}$:
\begin{equation}
    \sigma v_{\text{rel}} = \frac{1}{m_\chi \sqrt{s} }\text{Im}\mathcal{M}
\end{equation}
where $\sqrt{s}=2m_\chi+E$ is the total energy of the system, and for highly non-relativistic DM, $\sqrt{s}=2m_\chi$. Forward scattering amplitude for a pair of non-relativistic DM particles can be~\cite{Beneke:2022rjv}expressed as:
\begin{equation}
    i \mathcal{M} = \sum_{m,n}\int d^4 x \,
    \langle \chi \bar \chi|T\{ i J_m^\dagger(x) iJ_n(0)\} | \chi \bar \chi \rangle + \sum_k \left\langle \chi \bar \chi \middle|i T_k(0) \middle| \chi \bar \chi \right\rangle\,,
\label{eq:matrix_elem}
\end{equation}
where $T$ is the time-ordering operator, and $J_n$ are the production operator for the long-lived resonance states. The second term, described by $T_k$, accounts for non-resonant local interactions and is negligible compared to the first term due to suppression by $\Gamma_h/m_h$. For the simple benchmark model considered in the main text, the production operator is:
\begin{equation}
    J(x) = \sqrt{2 m_V}\,g_V (\chi^\dagger\gamma^{\mu} \chi V_{\mu})\,,
    \label{eq:production}
\end{equation}

In the non-relativistic limit,
\begin{equation}
    \begin{aligned}
    \chi^\dagger\gamma^{0} \chi&\approx\chi^\dagger\chi + O(v^2)\;,\\
    \chi^\dagger\gamma^{i} \chi&\approx\frac{1}{2m_\chi}\chi^\dagger \overleftrightarrow{D^i} \chi + O(v^3)\;.
    \label{eq:dirac}
    \end{aligned}
\end{equation}
Here, $\chi^\dagger\chi$ is the time component, and $\chi^\dagger \overleftrightarrow{D^i} \chi$ is the spatial component, which is suppressed by $v^2$ in the non-relativistic regime.

Substituting Eq. (\ref{eq:dirac}) into Eq. (\ref{eq:production}) and neglecting higher-order velocity terms yields:
\begin{equation}
    J(x) = \sqrt{2 m_V}\,g_V (\chi^\dagger \chi V_{\mu} \delta^{\mu0})\,,
\end{equation}

Since there is no long-range interaction between the DM and the resonance field, the matrix element in Eq. (\ref{eq:matrix_elem}) can be factorized into a product of two terms:
\begin{equation}
\begin{aligned}
    \int & d^4 x \,
    \langle \chi \bar \chi|T\{ i J^\dagger(x) iJ(0)\} | \chi \bar \chi \rangle \\
    =& i^2 2m_V g^2_V \int d^4 x \,\langle \chi \bar \chi|T\{[\chi^\dagger \chi]^\dagger(x)[\chi^\dagger \chi](0) \} | \chi \bar \chi \rangle \langle 0|T\{V^\dagger_0(x) V^0(0)\} | 0 \rangle \\
    =&i^2 2m_V g^2_V e^{iEt} \int d^4 x \, \langle \chi \bar \chi|[\chi^\dagger \chi]^\dagger(0)|0\rangle \langle0|[\chi^\dagger \chi](0) \}| \chi \bar \chi \rangle \langle 0|T\{ V^\dagger_0(x) V^0(0)\} | 0 \rangle \\
    =&i^2 2m_V g^2_V |\Psi_E(0)|^2\int d^4 x \, \langle 0|T\{ V^\dagger_0(x) V^0(0)\} | 0 \rangle e^{iEt}
\end{aligned}
\label{eq:matrix}
\end{equation}

The DM field wave function at $x=0$, $\Psi_E(0)$, is described by the potential non-relativistic DM (PNRDM) effective theory~\cite{Biondini:2021ccr,Jain:2020vgc}:
\begin{equation}
\mathcal{L}_{\rm PNRDM} = \chi^\dagger 
\left(i D_0 + \frac{\bf{\partial}^2}{2 m_\chi} \right) \chi 
- \int d^3 \mathbf{r} \, V(r) \left[\chi^\dagger \chi\right](t, \mathbf{x}) \left[\chi^\dagger \chi\right](t, \mathbf{x} + \mathbf{r})\,.
\label{eq:PNRDM}
\end{equation}
Here, $V(r)$ is the static Yukawa potential generated by the exchange of the light vector particle $A$:
\begin{align}
V(r) = - \frac{\alpha_A}{r} e^{-m_A r} \,,
\end{align}
This potential mediates the Sommerfeld effect, enhancing both DM annihilation and self-interactions. The Sommerfeld enhancement factor is defined as~\cite{Feng:2010zp}:
\begin{equation}
    S_{\text{SF}}(v)=\frac{|\Psi_E(0)|^2}{|\Psi_0(0)|^2}=|\Psi_E(0)|^2\,.
\end{equation}

The remaining resonance propagator in Eq. (\ref{eq:matrix}) is:
\begin{align}
    \int d^4 x \, \langle 0|T\{ V^\dagger_0(x) V^0(0)\} | 0 \rangle e^{iEt} = \frac{i\,\delta^{00}}{E-\delta M+i\frac{\Gamma_h}{2}}
    \label{eq:res-pro}
\end{align}

In summary, the scattering amplitude simplifies to:
\begin{equation}
    i \mathcal{M} = i^2 2m_V g^2_V S_{\text{SF}}(v)\frac{i}{E-\delta M+i\frac{\Gamma_h}{2}}\,,
    \label{eq:scat}
\end{equation}

Consequently, the DM annihilation cross section is:
\begin{equation}
    \sigma v_{\text{rel}}=\frac{g^2_V}{4m^2_{\chi}}\frac{m_{\chi}\Gamma_{V,\bar{l}l}/2}{(\frac{1}{4}m_{\chi}v^2 - \delta M)^2 + \Gamma^2_V/4}S_{\text{SF}}(v)\,,
\end{equation}

It is important to note that the resonance state $V$ does not decay exclusively into $l$ particles; the propagator term in the forward scattering amplitude includes the possibility of $V$ decaying into all allowed final states. Therefore, for the process $\chi\bar{\chi} \to V \to l\bar{l}$, the annihilation cross section should correspond to the probability of $V$ decaying into $l$ particles, i.e., $\Gamma_{V,\bar{l}l}$. Secondly, the constant prefactor in $\mathcal{M}$ is unimportant within a small range, as it can ultimately be absorbed into the coupling constant $g_l$ between the vector $V$ and the $l$ particles. As discussed in Sec.~\ref{sec:Self-interaction}, for any point in the DM self-interaction parameter space, one can always find a suitably small $g_l$ such that $\chi^2_{\text{anni}}$ is sufficiently small (compared to $\chi^2_{\text{SI}}$).

The DM self-scattering viscosity cross section is:
\begin{equation}
    \sigma=\frac{3}{2} \int \sin^2 \theta \frac{d\sigma}{d\Omega}d\Omega\,,
\end{equation}

In the non-relativistic limit, the differential scattering cross section is:
\begin{equation}
    \frac{d\sigma}{d\Omega} = \frac{1}{64\pi^2 s} \frac{|\vec{p}_f|}{|\vec{p}_i|} |M|^2\,,
\end{equation}

For identical particles, $|\vec{p}_i|=|\vec{p}_f|=m_{\chi}v_{\text{rel}}/2$. From Eq. (\ref{eq:scat}), we have:
\begin{equation}
    |\mathcal{M}|^2 = 4m^2_V g^4_V S^2_{\text{SF}}(v)\frac{1}{(\frac{1}{4}m_{\chi}v^2-\delta M)^2 + \Gamma^2_V/4}\,,
\end{equation}
where $g^4_V=64\pi^2\Gamma^2_{V,\chi\bar{\chi}}/(2m_V \delta M)$ (Eq.(\ref{eq:Gamma_xx})). Finally, considering that both $\chi$ and the resonance $V$ have spin degeneracies, the self-scattering cross section must be summed over all possible spins. Therefore, the total self-scattering cross section is:
\begin{equation}
    \sigma=\frac{8\pi \omega}{m_{\chi}\delta M}\cdot\frac{\Gamma^2_{V,\chi\bar{\chi}}/4}{(\frac{1}{4}m_{\chi}v^2-\delta M)^2 + \Gamma^2_V/4}\times S^2_{\text{SF}}(v)\,.
\end{equation}

\bibliography{ref}

\end{document}